\documentclass{ws-p8-50x6-00}

\usepackage{epsfig}
\usepackage{amssymb}
\usepackage{rotating}

\newcommand{\beq}{\begin{equation}}
\newcommand{\eeq}{\end{equation}}
\newcommand{\beqa}{\begin{eqnarray}}
\newcommand{\eeqa}{\end{eqnarray}}

\def\a{\alpha}
\def\g{\gamma}
\def\d{\delta}

\def\p{\pi}
\def\r{\rho}
\def\s{\sigma}
\def\f{\phi}
\def\W{\Omega}

\def\vs{\vskip}
\def\\{\hfill\break}

\def\ran{\rangle}
\def\lan{\langle}

\def\ifmath#1{\relax\ifmmode #1\else $#1$\fi}%
\def\rd{\ifmath{{\mathrm{d}}}}

\def\rt{\ifmath{{\mathrm{t}}}}
\def\rT{\ifmath{{\mathrm{T}}}}

\def\Re{\ifmath{{\mathrm{Re}}}}
\def\tot{\ifmath{{\mathrm{tot}}}}
\def\vec#1{{\mbox{\bf #1}}}
\def\lan{\langle}
\def\ran{\rangle}

\def\ifmath#1{\relax\ifmmode #1\else $#1$\fi}%

\begin{document}

\title{Correlations and Fluctuations \\
(an Introductory Review)}

\author{Wolfram Kittel}

\address{HEFIN, University of Nijmegen/NIKHEF, Toernooiveld 1, 6515 ED Nijmegen, NL\\
E-mail: wolfram@hef.kun.nl}


\maketitle

\abstracts{
Far too often are multiparticle final states studied and models tested on 
merely single-particle spectra and their integrals, the average multiplicities:
A multiparticle final state is a non-linear, complex system and the
essential information on the dynamics is contained in the particle
correlations and their integrals measuring the fluctuations!
}

\section{The formalism}
We start by defining symmetrized inclusive $q$-particle distributions
\beq
\r_q (p_1,\dots,p_q)= \s_\tot^{-1} \rd \s_q(p_1,\dots,p_q)/
\prod^q_1\rd p_q\ \ ,
\label{dat-1}
\eeq
where $\s_q(p_1,\dots,p_q)$ is the inclusive cross section for $q$
particles to be at $p_1,\dots,p_q$, irrespective of the presence and location
of any further particles, $p_i$ is the (four-) momentum of particle $i$ and
$\s_\tot$ is the total hadronic cross section of the collision under
study. For the case of identical particles, integration over an interval
$\W$ in $p$-space  yields
\vskip-1mm
\beq
 \int_\W \rd p_1 \dots \int_\W \rd p_q \r_q (p_1,\dots,p_q) =
\lan n(n-1)\dots (n-q+1)\ran \ ,
\label{dat-2}
\eeq
where $n$ is the multiplicity of identical particles within $\W$ in a 
given event and the angular brackets imply the average over the event ensemble.

Besides the interparticle {\it correlations}, the $\rho_q$ in general
contain ``trivial'' contributions from lower-order densities. 
We, therefore, consider a sequence of {\em cumulant
coorelation functions} $C_q(p_1,\dots,p_q)$ which vanish whenever one
of their arguments becomes statistically independent of the others:\cite{Mue71}
\vskip-2mm
\begin{eqnarray}
C_2(p_1,p_2)&=&\rho_2(p_1,p_2) -\rho_1(p_1)\rho_1(p_2)\ ,\nonumber\\
C_3(p_1,p_2,p_3)&=&\rho_3(p_1,p_2,p_3)
-\sum_{(3)}\rho_1(1)\rho_2(p_2,p_3)+2\rho_1(p_1)\rho_1(2p_)\rho_1(p_3)\ ,
\nonumber
\label{a:4b}
\end{eqnarray}
etc. Deviations of these functions from zero shall be addressed as 
{\it genuine} correlations. For recent discussions, it is important to note 
that not the $\r_q$ but the $C_q$ are additive in case of $N$ independent 
overlapping samples.

It is often convenient to normalize the functions $\rho_q$ and $C_q$ by the 
uncorrelated densities, 
\begin{eqnarray}
R_q(p_1,\dots,p_q) &=& \rho_q(p_q,\dots,p_q)/\rho_1(p_1)\ldots
\rho_1(p_q),\label{3.8}\\
K_q(p_1,\ldots,p_q)& =& C_q(p_1,\ldots,p_q)/\rho_1(p_1)\ldots
\rho_1(p_q).\label{3.9}
\end{eqnarray}
Important relations between the two sets of functions are:
\beq
R_2 = 1+K_2 \ ;\ \ R_3 =  1+3K_2+K_3
\label{dat-5}
\eeq
Because of the additivity of $C_q$, the $K_q$ will show a dilution according 
to $K^{(N)}_q = N^{(1-q)}K^{(1)}_q$ for $N$ independent overlapping sources.

Correlations have been studied extensively for $q=2$ and a strong positive 
correlation is found for $p_1\approx p_2$.\cite{wolf96} To be able to do that 
for $q\geq 3$, one must, however, work via the integrals Eq.(2), in limited 
phase-space cells.\cite{bialas}

In practical work, with limited statistics, it is almost always necessary 
to perform averages over more than a single phase-space cell. Let 
$\Omega_m$ be such a cell (e.g. a single rapidity interval of size $\delta y$)
and divide the phase-space volume $\Omega$ into $M$ non-overlapping equal 
cells $\Omega_m$ of size $\d\Omega=\Omega/M$. {\em Normalized cell-averaged 
factorial moments}~\cite{bialas} are then defined as
\def\hsm{\hskip-1mm}
\beq
F_q(\delta\Omega)\hsm\equiv\hsm\frac{1}{M}\hsm\sum^M_{m=1}\hskip-2mm
\frac{\int_{\delta\Omega} \rho_q(p_1,\ldots,p_q) \prod^q_{i=1}\hsm\rd p_i}
{\left(\int_{\d y} \r(p)\rd p\right)^q}
= \frac{1}{M}\hsm\sum_{m=1}^M\hsm
\frac{\langle n_m(n_m\hsm-\hsm1)\dots(n_m\hsm-\hsm q\hsm+\hsm1)\rangle}{\langle n_m\rangle^q}.
\label{dat-11}
\eeq
                 
Likewise, {\em cell-averaged normalized factorial cumulant moments} 
$K_q(\delta y)$ may be defined \cite{CaES91} by replacing the $\r_q$ by 
$C_q$ in (\ref{dat-11}). In analogy to Eq.~(\ref{dat-5}), they are related 
to the factorial moments by
\beq
F_2 = 1+K_2\ \ \ ;\ \ F_3 = 1+3K_2+K_3\ \ \ .\label{dr:48}
\eeq
Essential porperties are:

1.~The Poisson-noise suppression of the $F_q$, contrary to e.g. 
$\langle n^q\rangle/\langle n\rangle^q$.

2.~Independent emission leads to $F_q\equiv 1$ and 
$K_q\equiv 0$ for all $q$.

3.~The factorial moments resolve the high-$n_m$ tail of the multiplicity 
distribution and are, therefore, sensitive to high-density fluctuations.

A fruitful further development~\cite{Hen83} is the use 
of integrals of $\r_q$ or $C_q$ over a strip domain, rather than a 
sum of the box domains. One not only avoids unwanted side-effects 
such as splitting of density spikes, but also drastically increases the
statistical significance at a given resolution. It, furthermore, allows to 
work in terms of inter-particle distance measures, as the (invariant) 
four-momentum difference $Q=\sqrt{-(p_1-p_2)^2}$.

As shown in~\cite{bialas}, a ``smooth" (rapidity) distribution, which does 
not show any fluctuations except for the statistical ones, has the property 
of $F_q(\d y)$ being independent of the resolution $\d y$ in the limit 
$\d y\to  0$. On the other hand, if self-similar dynamical fluctuations exist, 
the $F_q$ obey the power law
\beq
F_q(\d y) \propto (\d y)^{-\f_q}\ , \ \ (\d y\to 0).
\label{dat-15}
\eeq
The powers $\f_q$ (slopes in a double-log plot) are related~\cite{LiBu89} 
to the anomalous dimensions $d_q=\f_q/(q-1)$,
a measure for the deviation from an integer dimension. Equation (\ref{dat-15})
is a scaling law since the ratio of the factorial moments at resolutions $L$ 
and $\ell$, $R = F_q(\ell)/F_q(L) = (L/\ell)^{\f_q}$, only depends on the 
ratio $L/\ell$, but not on $L$ and $\ell$, themselves.

The scaling behavior is related to the physics of fractals. Fractal behavior 
is indeed expected for multiparticle production both from QCD branching and 
from a QCD phase transition.

\section{The State of the Art on Power-Law Scaling (Intermittency)}
{\em 2.1 Existence of scaling}: The suggestion that the $F_q$ or $K_q$
might show power-law behavior has spurred a vigorous experimental search 
and (more or less) linear dependence of $\ln F_q$ on $-\ln\d \Omega$
has been found in all types of collisions.\cite{wolf96} 
Recent e$^+$e$^-$ results will be presented here by De Wolf.

\begin{figure}[tH]
\begin{minipage}[t]{5.8cm}
\epsfig{file=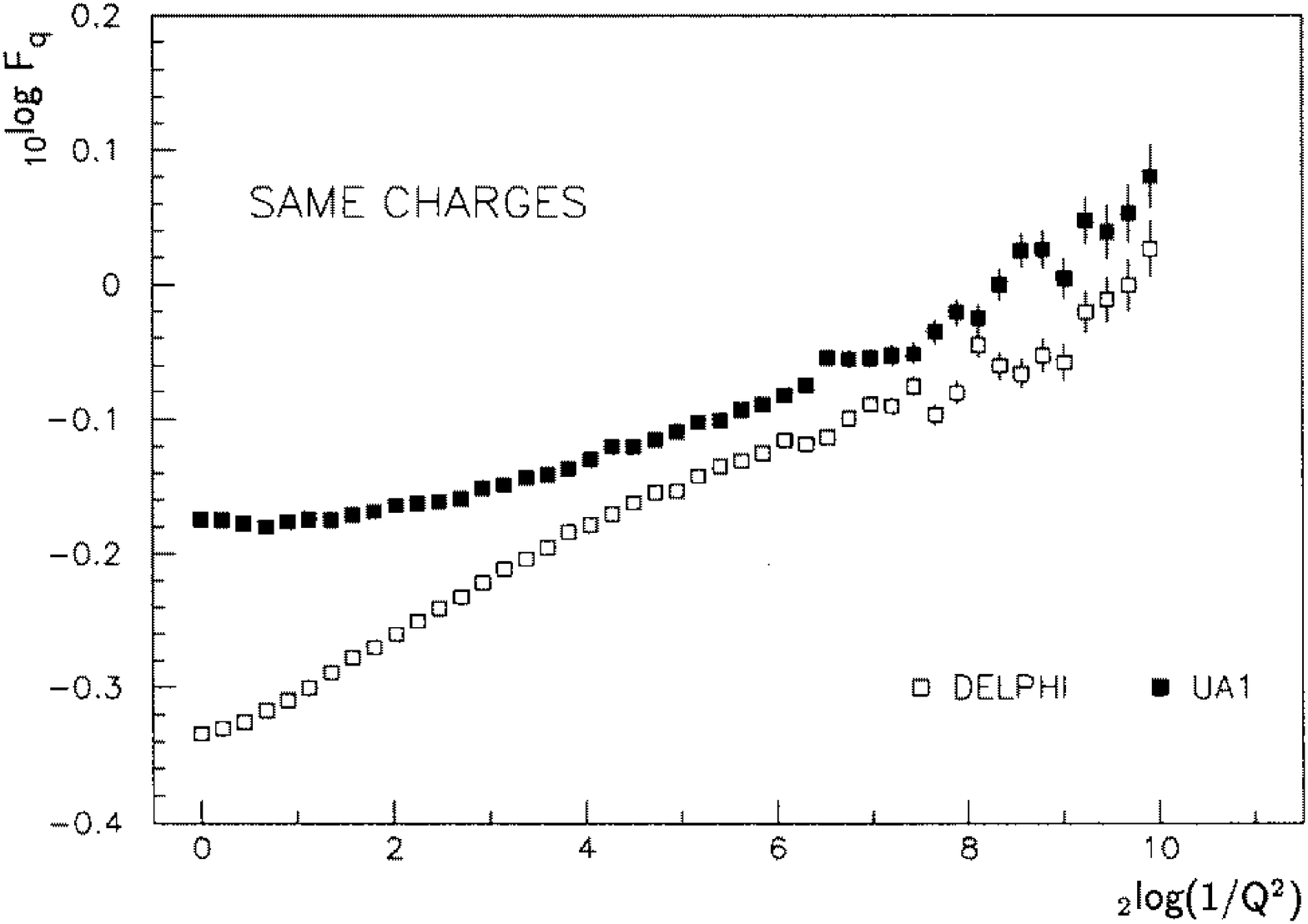,height=4.cm}
\end{minipage}
\begin{minipage}[t]{5.8cm}
\epsfig{file=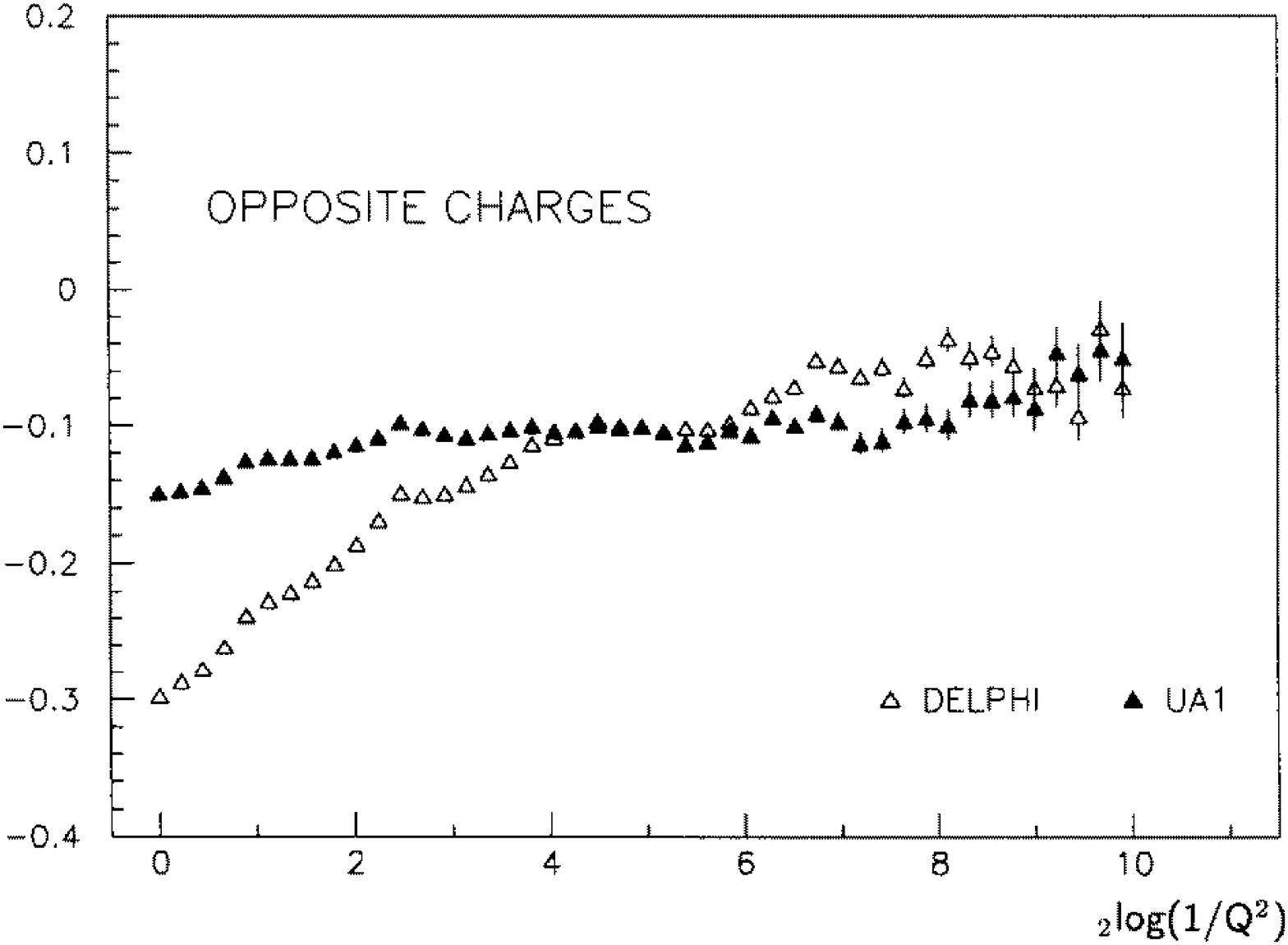,height=4.cm}
\end{minipage}

\caption{Comparison of density integrals for $q=2$ 
as a function of $_2\log(1/Q^2)$ for e$^+$e$^-$ (DELPHI) 
and hadron-hadron collisions (UA1).\protect\cite{Abreu94}}
\vs-5mm
\end{figure}

Of particular interest is a comparison of hadron-hadron to e$^+$e$^-$ results 
in terms of same and opposite charges of the particles involved. This 
comparison has been done for UA1 and DELPHI data in~\cite{Abreu94} and is 
shown in Fig.~1 for $q=2$. Important differences between UA1 
and DELPHI can be observed: 

For relatively large $Q^2(>0.03$ GeV$^2$), where Bose-Einstein effects do 
not play a major role, the e$^+$e$^-$ data increase much faster with 
increasing $_2\log(1/Q^2)$ than the hadron-hadron data. 
The increase in this $Q^2$ region is very similar for same and for 
opposite charges. At small $Q^2$, however, the e$^+$e$^-$ results 
approach the hadron-hadron results and the increase is faster for same
than for opposite charges. So, at least
two processes are responsible for the power-law behavior: Bose-Einstein
correlation at small $Q^2$ following the evolution of jets at larger $Q^2$. 
                  
As such, Bose-Einstein correlations from a static source are not
power-behaved. A power law is obtained \cite{Bial92} if i) the size of the
interaction region is allowed to fluctuate, and/or ii) the interaction region
itself is a self-similar object. 
 
\vskip 2mm\noindent
{\em  2.2 Multifractal or Monofractal}:
Fractals with all $d_q$ equal are called monofractals, those with $d_q$
depending on $q$ are called multifractals. A second-order phase transition
(e.g. from a QGP) would give rize to a monofratal behavior of particle
production, a cascade process (as e.g. QCD branching) leads to a
multifractal behavior. Even though the $d_q$ and their $q$ dependence
are found to be largest in e$^+$e$^-$ and smallest in heavy-ion collisions,
so far all types of collisions (including AA collisions) exhibit multifractal 
behavior. We shall hear more about monofractal behavior from N. Antoniou.

\vskip 2mm\noindent
{\em 2.3 Self-similar or Self-affine Fractal}:
If the power law (\ref{dat-15}) holds when $p$-space is partitioned by the same
factor in different directions, the fractal is called {\it self-similar}. If, 
on the other hand, (\ref{dat-15}) holds and only holds when space is 
partitioned by different factors in different directions, the corresponding 
fractal is called {\it self-affine}.\cite{Wu93}

Self-affinity can be characterized
by the Hurst exponents,\cite{Wu93}
$H_{ij}=\ln M_i/\ln M_j \ \ \ (0\leq H_{ij}\leq 1)$
with $M_i$ ($i=1,2,3;\ \ M_1\leq M_2\leq M_3$) being the partition numbers
in the phase-space variables $p_i$.
The $H_{ij}$ can be obtained from the 
saturation of the one-dimensional ln$F_2(\d p_i)$ distributions.

For $hh$ collisions, the $H_{ij}$ for longitudinal-transverse
combinations is indeed smaller than unity for NA22 
\cite{NA22-98} and NA27 \cite{NA27-2}, while it is consistent with 
unity within the transverse plane (see Wang Shaoshun for hh and Chen Chang 
for first results on e$^+$e$^-$ collisions)

\vskip2mm\noindent
{\em 2.4 }: 
QCD + LPHD have been quite successful~\cite{drem01} in the description of
single-particle $\xi$ distributions and the energy evolution of its
maximum position $\xi^*$ and integral $n$. According to Eqs.~(\ref{dat-1}) 
and (\ref{dat-2}), these correspond to $q=1$ only and, therefore, neglect 
any vital information on particle correlations.\\
a. {\em High-Order Correlations:}
The ratio $H_q=K_q/F_q$ reflects the genuine $q$-particle 
correlation relative to the $q$-particle density. They have been 
calculated for gluons at different orders of QCD.\cite{dremin2}
For NNLLA,
$H_q$ decreases to a negative first minimum for $q=5$ and shows 
quasi-oscillation around 0 for $q>5$.
Assuming the validity of LPHD, such a behavior is also expected for the 
charged particles.
For e$^+$e$^-\to \mathrm{hadrons}$ at $\sqrt{s}=M_{\mathrm{Z}}$,
SLD~\cite{SLD} indeed observed such a behavior.
However, the same behaviour is observed 
in hh collisions between 20 and 900 GeV \cite{drem94} and
even in hA and AA collisions.\cite{capel97}

An L3 analysis of sub-jet multiplicities \cite{l3hq}
reveals that this behavior appears only for 
energy scales $\lesssim 100$ MeV, 
far away from the perturbative region. Furthermore, similar oscillatory
behavior is observed for a large variety of MC models, including independent
fragmentation.
One is tempted to conclude, that at present energies, the oscillations 
of $H_q$ are unrelated to the behavior predicted by the NNLLA calculations.
They may be related to the energy conservation first approximated in
NNLLA, but what do we have to think of a theory approximating 
energy conservation only in such a high order of perturbative expansion?

The oscillations are reproduced by a superposition of 3 negative binomials
\cite{giovan} corresponding to two-jet, mercedes-like three-jet and
intermediate events, respectively.\cite{l3hq} So, the oscillations
originate from an interplay between hard (jet topology) and soft 
(fragmentation) phenomena.

\vskip 2mm\noindent
b) {\em Local fluctuations:}
In perturbative QCD, the fractal structure of jets follows from
parton branching and the $\f_q$ are directly 
related~\cite{Gust91} to the anomalous dimension 
$\g_0 = (6\a_s/\p)^{1/2}$, at lowest order via
$d_q = D-\gamma_0(q+1)/q$,
where $D$ is the topological dimension of the analysis and the
second term corresponds to the R\'enyi dimensions $D_q$ of the fractal.

Analytical QCD predictions have indeed been given
for fluctuations in angular intervals.\cite{Gust91} 
Assuming LPHD once again, these  predictions can be 
compared to experimental data.
DELPHI \cite{del}, L3 \cite{Acci98} and ZEUS \cite{chek} show that 
the predictions disagree with the data.
Energy-conservation effects are indeed sizeable.\cite{cons} 
Recent ZEUS results will be presented here by L. Zawiejski and the
state of understanding will be summarized by W. Ochs.

\vskip2mm\noindent
{\em 2.5 Event-to-Event Fluctuations, Erraticity, Voids and Entropy}:
The $F_q$ as defined in (\ref{dat-11}) contain averages over all bins within 
one event and over the whole event sample. Event-to-event fluctuations lost in
this procedure can be quantified \cite{Cao} by defining the $F^e_q$ per event
and determining moments of the distribution $P(F^e_q)$ as 
$C_{q,p}=\lan (F^e_q)^p\ran/\lan F^e_q\ran^p$.
However, in an analysis of the influence of statistical
fluctuations,\cite{Fu}
 it turned out that these completely dominate, even in heavy-ion collision.

When the event multiplicity is low, the gaps between neighboring particles
carry more information than multiplicity spikes. The method has
therefore been extended to measure the rapidity gaps (voids).\cite{Hwa00}
Experimentally the erraticity measures have been studied on NA27 \cite{Wang01}
(see Wang Shaoshun). 

A method most sensitive to the maximum of the probability distribution is a 
measure of entropy suggested in \cite{Bia99}.

Furthermore, results based on event-to-event $p_\rt$ fluctuations
will be presented here by Bai Yuting and T. Sugitate.

\section{Bose-Einstein Correlations}

{\em 3.1 Alternative Views}:
As can be seen on Fig.~1, BEC play a role at small $Q$, both in hh and 
e$^+$e$^-$ collisions. Conventionally, they are studied within intensity 
interferometry.\cite{Hanb54} Basic ingredients are incoherent (chaotic) boson 
emission from a spherically symmetric static source and production amplitudes 
independent of momentum. Undere these conditions,
\beqa
R_2 &\equiv& 1+K_2 = 1+|\tilde \r(Q)|^2 \\
R_3 &=& 1+|\tilde\r_{12}|^2+|\tilde\r_{23}|^2 + |\tilde\r_{31}|^2+
2\Re(\tilde\r_{12}\tilde\r_{23}\tilde\r_{31})\ ,
\eeqa
where $\tilde\r(Q)$ is the Fourier transform of the space-time density of
boson emitters. L. Gurvits will show us, however, how far 
astrophysics has developed away from intensity to amplitude interferometry,
in the meantime.

A formalism particularly handy for the fully-dimensional treatment of 
a dynamical 
emitter is the so-called Wigner-function formalism.\cite{Gyu79}
This is based on the emission function $S(x,p)$, a covariant Wigner-transform
of the source density matrix. $S(x,p)$ can be interpreted as a
quantum-mechanical analogue of the classical probability that a boson is 
produced at a given space-time point $x=(t,\vec r)$ with a given 
momentum-energy $p=(E,\vec p)$ (see T. Cs\"org\H{o}).

Alternatively, BEC are introduced
into string models.\cite{120} An ordering in 
space-time exists for the hadron momenta within a string.
Bosons close in phase space are nearby in space-time and 
the length scale measured by BEC is not
the length of the string, but the distance in boson-production
points for which the momentum distributions still overlap (B. Andersson).

\vskip2mm\noindent
{\em 3.2 Recent experimental results}:
Bose-Einstein correlations are by now a well established effect in the
hadronic final states of Z and W decay (intra-W BEC).\cite{acto91} 
The important 
question is that of BEC between pions each originating from a different W in 
fully hadronic W$^+$W$^-$ final states (inter-W BEC). If existent, such a 
correlation could, on the one hand, cause a bias in the mass determination of 
the W. On the other, it could serve as a pion-interferometry laboratory for 
the measurement of the space-time development. The recent status of the 
search for inter-W BEC will be covered by J. van Dalen.

Detailed knowledge of intra-W BEC would be necessary to understand
inter-W BEC. Limited statistics prevents that, but a large amount
of high statistics information has recently been obtained for the Z and
other reactions. For shortness, I am forced to refer to recent Zakopane 
lectures \cite{kittel} on this topic and to limit myself to just
listing the main results:
\vskip 1mm
{\em Bose-Einstein correlations exist} also in e$^+$e$^-$ collisions and
DIS, with radius parameters independent of $\sqrt s$ and $Q^2$, respectively
(E. De Wolf and L. Zawiejski).
\vskip 1mm
{\em The region of homogeneity} is elongated along the event axis.\cite{L3E}
This is not reproduced by JETSET + LUBOEI, but expected from\cite{120}.
The short range of this region is evidence for a strong momentum-position 
correlation.
\vskip1mm
{\em The correlator $K_2$ is steeper than Gaussian}.\cite{agab93,acto91}
This is contrary to most of the assumptions and interpretations in the 
literature, but, again expected from \cite{120}.
\vskip 1mm
{\em All radius parameters show a $1/\sqrt m_\rT$ dependence} in all
types of reactions.\cite{NA44-2} This may be explainable already from the 
Heisenberg principle,\cite{alex99} but is expected from an inside-outside 
cascade with transverse flow.\cite{biazal99} More results from RHIC here 
from T. Sugitate.
\vskip 1mm
{\em In hh and AA collisions, there is evidence for dilution} of $K_2$ due to 
independence of overlapping mechanisms.\cite{busmat} Search for a similar 
dilution due to overlap of string pieces in gluon jets will be presented by 
N. van Remortel.
\vskip 1mm
{\em The pion emission function} in space-time has been extracted
for hh and AA collisions.\cite{agab97} In the transverse plane, it is 
Gaussian for AA collisions but ring-shaped (!) for hh collisions. Study of 
the Z is underway (T. Cs\"org\H{o}).
\vskip 1mm
{\em For $\p^0\p^0$}, according to the string model, the radius should be
smaller than for charged pions. Wes Metzger will present the experimental
evidence.
\vskip1mm
{\em Genuine three-particle correlations} contain additional information
on a phase or, alternatively, on dilution of $K_q$ proportional to $N^{(1-q)}$
for $N$ independent sources. W. Metzger will give recent results on the Z.
\vskip1mm
{\em Modelling} these observations in a Monte Carlo code is a major challenge
(see G. Wilk, K. Fia\l kowski).

\section{Conclusion}

We need a model containing parton branching and energy conservation plus
soft fragmentation to reproduce the fractal behavior of particle production.
JETSET and ARIADNE are doing surprisingly well there,
HERWIG a bit less.
We further need the model to do away with old prejudice from conventional
pion interferometry.
A large amount of new information is (becoming) available on BEC at the Z.
Models neglecting this information I would consider a loss of time.


\begin{thebibliography}{99}
\bibitem{Mue71}
A.H. Mueller, {\em Phys. Rev.} {\bf D4}, 150 (1971).
 
\bibitem{wolf96}
E.A. De Wolf {\em et al}, {\em Phys. Rep.} {\bf 270}, 1 (1996).

\bibitem{bialas}
A. Bia\l as, R. Peschanski, {\em Nucl. Phys.} {\bf B273}, 703 (1986); 
ibid. {\bf B308}, 857 (1988).

 \bibitem{CaES91}
P. Carruthers {\em et al}, {\em Phys. Lett.} {\bf B 54}, 258 (1991).

\bibitem{Hen83}
P. Lipa {\em et al}, {\em Phys. Lett.} {\bf B285}, 300 (1992). 
 
\bibitem{LiBu89}
P. Lipa, B. Buschbeck, {\em Phys. Lett.} {\bf B223}, 465 (1989);
R. Hwa, {\em Phys. Rev.} {\bf D41}, 1456 (1990).
 
\bibitem{Abreu94} 
F. Mandl, B. Buschbeck in {\em Proc. Cracow Workshop on Multiparticle
Production}, eds. A. Bia\l as {\em et al} (World Scientific, 
Singapore, 1994) p.1.

\bibitem{Bial92}
A. Bia\l as,  {\em Acta Phys. Pol.} {\bf B23}, 561 (1992).

\bibitem{Gust91}
G. Gustafson, A. Nilsson, {\em Z. Phys.} {\bf C52}, 533 (1991); 
W. Ochs, J. Wosiek, {\em Phys. Lett.} {\bf B289}, 159 (1992);
ibid. {\bf 305}, 144 (1993); {\em Z. Phys.} {\bf C68}, 269 (1995);
Y.L. Dokshitzer, I.M. Dremin, {\em Nucl. Phys.} {\bf B402}, 139 (1993);
Ph. Brax {\em et al}, {\em Z. Phys.} {\bf C62}, 649 (1994).

\bibitem{ochs}
W. Ochs, {\em Phys. Lett.} {\bf B247}, 101 (1990); {\em Z. Phys.} {\bf C50},
339 (1991).

\bibitem{Wu93}
Y.F. Wu, L.S. Liu, {\em Phys. Rev. Lett.} {\bf 70}, 3197 (1993);
Y.F. Wu {\em et al}, {\em Phys. Rev.} {\bf D51}, 6576(1995);
Liu Feng {\em et al}, {\em Phys. Rev.} {\bf D59}, 114020 (1999).

\bibitem{NA22-98}
N.M. Agababyan {\em et al} (NA22), {\em Phys. Lett.} {\bf B382}, 305 (1996);
\\ ibid. {\bf B431}, 451 (1998).

\bibitem{NA27-2}
Wang Shaoshun {\em et al}, {\em Phys. Lett.} {\bf B410},
323 (1987).

\bibitem{drem01}
I.M. Dremin, J.W. Gary, {\em Phys. Rep.} {\bf 349}, 301 (2001).

\bibitem{dremin2}
I.M.~Dremin, {\em Phys. Lett.} {\bf B313}, 209 (1993);
I.M.~Dremin, V.A.~Nechita\v{\i}lo, {\em JETP Lett.} {\bf 58}, 811 (1993);
I.M.~Dremin, {\em Physics-Uspekhi} {\bf 37}, 715 (1994).

\bibitem{SLD}
K.~Abe {\em et al} (SLD Collab.), {\em Phys. Lett.} {\bf B371}, 149 (1996).

\bibitem{drem94}
I.M. Dremin {\em et al}, {\em Phys. Lett.} {\bf B336}, 199 (1994);
N.~Nakajima {\em et al}, {\em Phys. Rev.} {\bf D54}, 4333 (1996); 
Wang Shaoshun {\em et al}, {\em Phys. Rev.} {\bf D56}, 1668 (1997).

\bibitem{capel97}
A. Capella {\em et al}, {\em Z. Phys.} {\bf C75}, 89 (1997); 
I.M. Dremin {\em et al}, {\em Phys. Lett.} {\bf B403}, 149 (1997).
 
\bibitem{l3hq}
P. Achard {\em et al} (L3), subm. to Phys. Lett. B.

\bibitem{giovan}
A. Giovannini {\em et al}, {\em Phys. Lett.} {\bf B374},
231 (1996); {\bf B388}, 639 (1996).

\bibitem{del}
P. Abreu {\em et al} (DELPHI Collab.), {\em Phys. Lett.} {\bf B457},
368 (1999).

\bibitem{Acci98}
M. Acciari {\em et al} (L3 Collab.), {\em Phys. Lett.} {\bf B428}, 186 (1998).

\bibitem{chek}
S. Chekanov {\em et al} (ZEUS), {\em Phys. Lett.} {\bf B510}, 36 (2001).

\bibitem{cons}
J.-L. Meunier, R. Peschanski, {\em Z. Phys.} {\bf C72}, 647 (1996).

\bibitem{Cao}
Z. Cao, R.C. Hwa, {\em Phys. Rev. Lett.} {\bf 75}, 1268 (1995); 
{\em Phys. Rev.} {\bf D53}, 6608 (1996); ibid. {\bf D54}, 6674 (1996); 
R.C. Hwa, {\em Acta Phys. Pol.} {\bf B27}, 1789 (1996).

\bibitem{Fu}
Liu Lianshou {\em et al}, {\em Science in China} {\bf A30}, 432
(2000); Fu Jinghua {\em et al}, {\em Phys. Lett.} {\bf B472},
161 (2000); Liu Fuming {\em et al}, {\em Phys. Lett.} {\bf B516} (2001) 293.

\bibitem{Hwa00}
R.C. Hwa, W. Zhang, {\em Phys. Rev.} {\bf D62},  0140003 (2000);
R.C. Hwa, Y. Wu, {\em Phys. Rev.} {\bf D60} 097501 (1999).

\bibitem{Wang01}
Wang Shaoshun, Wu Chong, {\em Phys. Lett.} {\bf B505}, 43 (2001).

\bibitem{Bia99}
A. Bia\l as {\em et al}, {\em Acta Phys. Pol.} {\bf B30}, 107 (1999);
A. Bia\l as, W. Czy\.z, {\em Phys. Rev.} {\bf D61} 074021 (2000);
{\em Acta Phys. Pol.} {\bf B31}, 687 (2000).

\bibitem{Hanb54}
R. Hanbury Brown, R.Q. Twiss, {\em Phil. Mag.} {\bf 45}, 663 (1954).

\bibitem{Gyu79}
H. Gyulassy {\em et al}, {\em Phys. Rev.} {\bf C20}, 2267 (1979);
S. Pratt {\em et al}, {\em Phys. Rev.} {\bf C42}, 2646 (1990);
S. Chapman, U. Heinz, {\em Phys. Lett.} {\bf B340}, 250 (1994).

\bibitem{120}
B. Andersson, W. Hofmann, {\em Phys. Lett.} {\bf B169}, 364 (1986);
B. Andersson, M. Ringn\'er, {\em Phys. Lett.} {\bf B421}, 283 (1998) and
  {\em Nucl. Phys.} {\bf B513}, 627 (1998);
\v{S}. Todorova-Nov\'a, J. Rame\v{s}, Strasbourg preprint IReS97-29.

\bibitem{acto91}
P.D. Acton {\em et al} (OPAL), {\em Phys. Lett.} {\bf B267}, 143 (1991);
P. Abreu {\em et al} (DELPHI), {\em Phys. Lett.} {\bf B286}, 201 (1992); 
{\em Z. Phys.} {\bf C63}, 17 (1994);
D. Decamp {\em et al} (ALEPH), {\em Z. Phys.} {\bf C54}, 75 (1992);
M. Acciari {\em et al} (L3), {\em Phys. Lett.} {\bf B493}, 233 (2000).

\bibitem{kittel}
W. Kittel, hep-ph/0110088, {\em Acta Phys. Pol.} (to be published).

\bibitem{L3E}
M. Acciarri {\em et al} (L3), {\em Phys. Lett.} {\bf B458}, 517 (1999);
P. Abreu {\em et al} (DELPHI), {\em Phys. Lett.} {\bf B471}, 460 (2000);
G. Abbiendi {\em et al} (OPAL), {\em Z. Phys.} {\bf C16}, 423 (2000).

\bibitem{agab93}
N.M. Agababyan {\em et al} (NA22), {\em Z. Phys.} {\bf C59}, 405 (1993);
N. Neumeister {\em et al} (UA1), {\em Z. Phys.} {\bf C60}, 633 (1993).

\bibitem{NA44-2}
G. Bearden {\em et al} (NA44), {\em Phys. Rev.} {\bf C58}, 1656 (1998);
B. L\"orstad, O.G. Smirnova, {\em Proc. 7th Int. Workshop on Multiparticle 
Production}, Nijmegen, eds. R.C. Hwa et al. (WSPC, Singapore, 1997) p.42.

\bibitem{alex99}
G. Alexander {\em et al}, {\em Phys. Lett.} {\bf B452}, 159 (1999);
G. Alexander, {\em Phys. Lett.} {\bf B506}, 45 (2001).

\bibitem{biazal99}
A. Bia\l as, K. Zalewski, {\em Acta Phys. Pol.} {\bf B30}, 359 (1999).

\bibitem{busmat}
B. Buschbeck {\em et al}, {\em Phys. Lett.} {\bf B481}, 187
(2000); {\em Nucl. Phys.} {\bf B} (Proc. Suppl.) {\bf 92}, 235 (2001).

\bibitem{agab97}
N.M. Agababyan {\em et al} (NA22), {\em Phys. Lett.} {\bf B422}, 359 (1998);
A. Ster {\em et al}, {\em Nucl. Phys.} {\bf A661}, 419 (1999);
R. Hakobyan in {\em Proc. XXX Int. Symp. on Multiparticle Dynamics}, eds.
R. Cs\"org\H{o} {\em et al} (WSPC, Singapore, 2001) p.331.

\end{thebibliography}
\end{document}
